\documentclass[aps,prl,twocolumn,superscriptaddress]{revtex4-2}

\usepackage{graphicx}
\usepackage{epsfig}
\usepackage{dcolumn}
\usepackage{bm}
\usepackage{overpic}
\usepackage{color}
\usepackage{multirow}
\usepackage{subfigure}
\usepackage{lineno}
\usepackage{amsmath}
\usepackage{url}

\usepackage{xcolor}

\usepackage{hyperref}
\hypersetup{
colorlinks=true,
linkcolor=blue,
filecolor=blue,
urlcolor=blue,
citecolor=blue,
pdftitle={LaTeX Example},
pdfpagemode=FullScreen,
}

\newcommand{\BESIIIorcid}[1]{\href{https://orcid.org/#1}{\hspace*{0.1em}\raisebox{-0.45ex}{\includegraphics[width=1em]{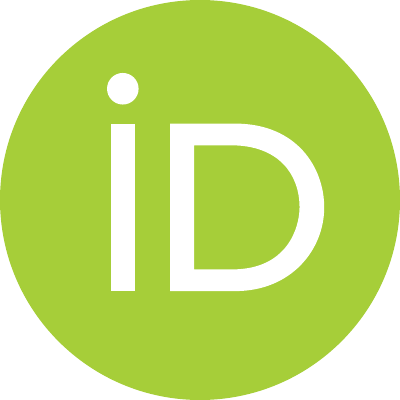}}}}

\begin{document}


\title{First Measurement of the Relative Phase between Proton Psionic Form Factors}
\author{
M.~Ablikim$^{1}$\BESIIIorcid{0000-0002-3935-619X},
M.~N.~Achasov$^{4,c}$\BESIIIorcid{0000-0002-9400-8622},
P.~Adlarson$^{81}$\BESIIIorcid{0000-0001-6280-3851},
X.~C.~Ai$^{87}$\BESIIIorcid{0000-0003-3856-2415},
C.~S.~Akondi$^{31A,31B}$\BESIIIorcid{0000-0001-6303-5217},
R.~Aliberti$^{39}$\BESIIIorcid{0000-0003-3500-4012},
A.~Amoroso$^{80A,80C}$\BESIIIorcid{0000-0002-3095-8610},
Q.~An$^{77,64,\dagger}$,
Y.~H.~An$^{87}$\BESIIIorcid{0009-0008-3419-0849},
Y.~Bai$^{62}$\BESIIIorcid{0000-0001-6593-5665}
% ...
\\
\vspace{0.2cm}
(BESIII Collaboration)\\
\vspace{0.2cm} {\it
$^{1}$ Institute of High Energy Physics, Beijing 100049, People's Republic of China\\
% ...
$^{4}$ Budker Institute of Nuclear Physics SB RAS (BINP), Novosibirsk 630090, Russia\\
% ...
$^{31}$ INFN Sezione di Ferrara, (A)INFN Sezione di Ferrara, I-44122, Ferrara, Italy; (B)University of Ferrara, I-44122, Ferrara, Italy\\
% ...
$^{39}$ Johannes Gutenberg University of Mainz, Johann-Joachim-Becher-Weg 45, D-55099 Mainz, Germany\\
% ...
$^{62}$ Southeast University, Nanjing 211100, People's Republic of China\\
% ...
$^{64}$ State Key Laboratory of Particle Detection and Electronics, Beijing 100049, Hefei 230026, People's Republic of China\\
% ...
$^{77}$ University of Science and Technology of China, Hefei 230026, People's Republic of China\\
% ...
$^{80}$ University of Turin and INFN, (A)University of Turin, I-10125, Turin, Italy; (B)University of Eastern Piedmont, I-15121, Alessandria, Italy; (C)INFN, I-10125, Turin, Italy\\
$^{81}$ Uppsala University, Box 516, SE-75120 Uppsala, Sweden\\
% ...
$^{87}$ Zhengzhou University, Zhengzhou 450001, People's Republic of China\\

\vspace{0.2cm}
$^{\dagger}$ Deceased\\
% ...
$^{c}$ Also at the Novosibirsk State University, Novosibirsk, 630090, Russia\\
% ...
}
}
\date{\today}

\begin{abstract}

The relative phase between the time-like form factors of the proton is a
crucial observable for a complete understanding of its internal
structure, yet it has remained unmeasured due to the formidable
experimental challenge of determining the final-state polarization or
having available polarized beams. With a novel technique that measures
polarization via secondary scattering on spectrometer material, we
use $10.09\times10^{9}$ $J/\psi$ events collected at BESIII to analyze
the reaction $e^+e^-\rightarrow J/\psi \rightarrow p\bar{p}$. This
allows the first determination of the sine of the relative phase
between the proton psionic form factors,
$\sin\Delta\Phi=-0.20\pm0.34_{\textrm{stat}}\pm0.11_{\textrm{syst}}$. This
result provides the first direct insight into the complex dynamics of
proton formation, and offers valuable new information to constrain
theoretical models of nucleon structure.

\end{abstract}

\maketitle

The electromagnetic (EM) structure of the proton is encoded in its
form factors (FFs). These describe the proton's interaction with a
virtual photon in two distinct regimes: 1) the space-like (SL) region,
with the four-momentum transfer squared $q^{2}<0$, probed in
electron--proton scattering, and 2) the time-like (TL) region
($q^2>0$), accessed in electron--positron annihilation into
proton--antiproton pairs (or the reverse process).  Historically,
elastic scattering experiments revealed that the proton is not a
point-like particle but instead has an extended spatial distribution of
charge and magnetization, parameterized by the electric $G_{E}$ and
magnetic $G_{M}$ FFs,
respectively~\cite{hofs1955Apr,hofs1956May,hofs1956Sep}. These
quantities are fundamental to understanding the nucleon structure and
have been precisely measured in the SL region using the Rosenbluth
separation~\cite{rosenbluth1950} and the polarization transfer
techniques~\cite{Akhiezer:1973xbf,galynskii2024}. In recent years, the
moduli of proton electromagnetic FFs (EMFFs) $|G_{E}|$ and $|G_{M}|$
in the TL region have been measured with high precision across a broad
$q^{2}$ range by experiments such as
$BABAR$~\cite{babarproton01,babarproton02},
BESIII~\cite{bes3proton2015,bes3proton2019,be3proton2020,bes3proton2021}
and CMD-3~\cite{AKHMETSHIN2016634}.

\begin{figure}[htbp]
\centering
\includegraphics[width=0.5\textwidth]{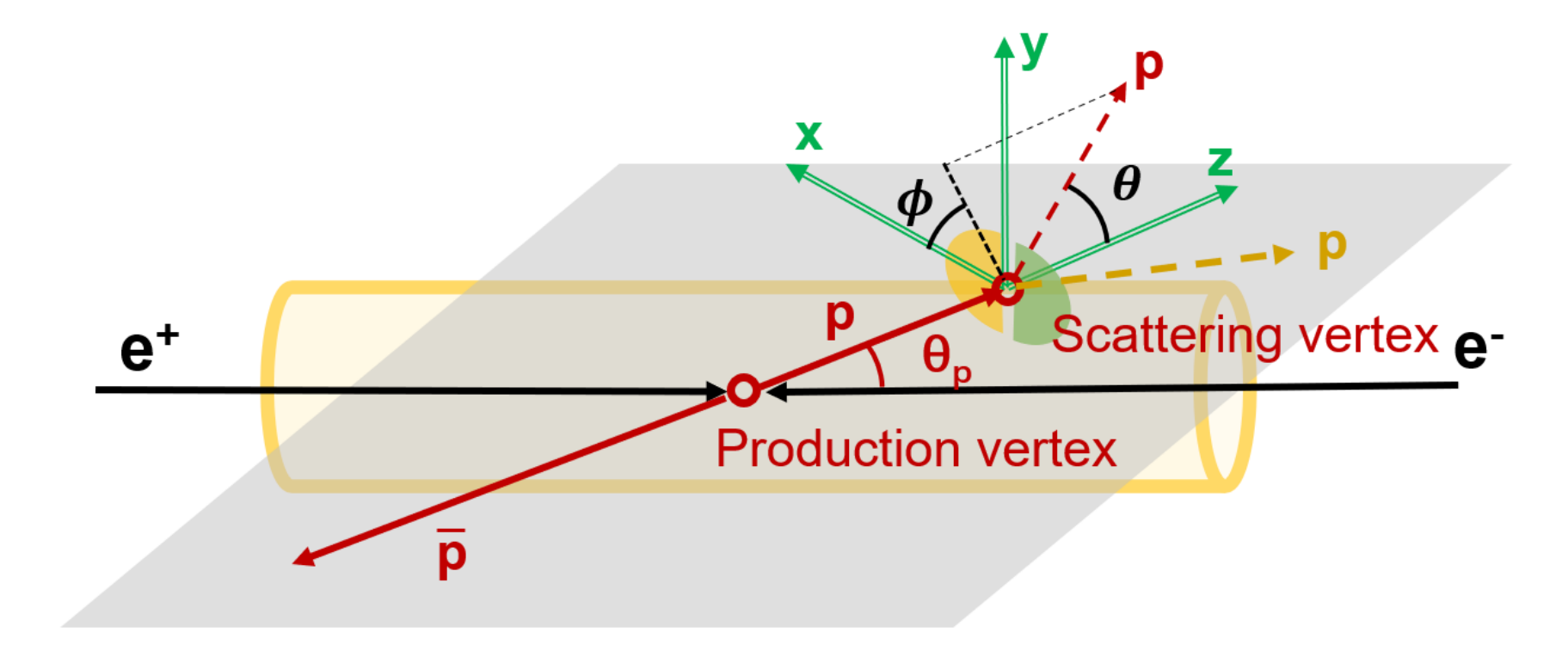}
\caption{Schematic of the reaction $e^{+}e^{-}\to p\bar{p}$ in the
$e^+e^-$ c.m.\ frame, which also defines the scattering plane. The
yellow cylinder represents the scattering layer. The proton polar
angle $\theta_p$ is defined between the proton momentum and the
positron momentum, which also defines the scattering plane. At the
scattering vertex, $\theta$ and $\phi$ are defined in the helicity
frame of proton, which is indicated by the $xyz$ coordinate system
marked with double-line arrows.}
\label{fig:illustration_coordinate}
\end{figure}

A complete determination of the spin-$\frac{1}{2}$ baryon's EMFFs in the
TL region can be achieved through one of two approaches. The
single-polarization method employs a longitudinally polarized lepton
beam while measuring the polarization of the outgoing
baryon~\cite{Dubnickova:1992ii,tomasi2005}. Alternatively, the
double-polarization method uses an unpolarized initial state but
requires a complete measurement of the polarizations and spin
correlations of the final-state baryon--antibaryon
pair~\cite{Dubnickova:1992ii,tomasi2005}.  Hyperons such as $\Lambda$
and $\Sigma$ benefit from their self-analyzing polarization
properties, allowing for a full determination of their EMFFs via the
double polarization
method~\cite{perotti2019,PhysRevLett.123.122003,PhysRevLett.132.081904}. For
nucleons, however, measuring the polarization of the outgoing particles
poses significant experimental challenges since it relies on secondary
scattering. In $e^{+}e^{-}$ annihilation, single-polarization studies
provide three nucleon polarization observables along axes forming a right-handed
coordinate system as shown in Fig.~\ref{fig:illustration_coordinate}:
1) longitudinal $\mathcal{P}_z$ along the outgoing baryon's momentum,
2) sideways $\mathcal{P}_x$ perpendicular to $\mathcal{P}_z$ within
the scattering plane, and 3) normal $\mathcal{P}_y$ perpendicular to
the scattering plane. While measuring
$\mathcal{P}_z$ and $\mathcal{P}_x$ requires a polarized initial
lepton beam, $\mathcal{P}_y$ can be measured with an unpolarized beam.

This polarization framework is also necessary for analyzing
baryon-antibaryon pair production in charmonium decays ($J/\psi$ or
$\psi(2S)$). In this process, the EMFFs are replaced by the psionic
FFs ($G_{E,M}^\psi$)~\cite{PhysRevD.77.056001,FALDT201716,xucao2024},
which represent the complex amplitudes of the charmonium decay. Given
the unpolarized initial lepton beams at BESIII, the normal
polarization $\mathcal{P}_y$ is the only accessible observable, from
which one can determine the sine of the relative
phase~\cite{Dubnickova:1992ii,PhysRevD.69.054022,xucao2024},
\begin{equation} \mathcal{P}_y =
\frac{\sqrt{1-\alpha^2_{J/\psi}}\cos\theta_{p}\sin\theta_{p}\sin\Delta\Phi}{1+\alpha_{J/\psi}\cos^2\theta_{p}},
\label{eq:py_angular_dependent}
\end{equation} 
where $\theta_{p}$ is the proton polar angle in the $e^{+}e^{-}$
center-of-mass (c.m.) frame, and $\Delta\Phi$ is the relative phase of
the complex EMFFs (or psionic FFs) in the TL region, $\alpha_{J/\psi}
=
\frac{\tau-|G_{E}^{\psi}/G_{M}^{\psi}|^{2}}{\tau+|G_{E}^{\psi}/G_{M}^{\psi}|^{2}}$
is the decay parameter of $J/\psi\to p\bar{p}$~\cite{BESIII:2012ion},
and $\tau = \frac{q^2}{4m^2_p}$, with $m_{p}$ being the proton
mass.

A measurement of the proton polarization usually utilizes the
spin-dependent cross section for the proton--proton $pp$ or
proton--carbon $p\textrm{C}$ elastic scattering given their large
analyzing powers. In the case of an unpolarized target, the
differential cross section for a given polar angle $\theta$ in the
helicity frame is expressed as~\cite{PhysRevLett.90.142301,
Bystricky:1976jr} \begin{equation}
    \frac{{\rm d}\sigma}{{\rm d}\phi {\rm d}\!\cos\theta} =  \frac{1}{2\pi}\frac{{\rm d}\sigma_0}{ {\rm d}\!\cos\theta}\left[1+\mathcal{P}_yA_N\!(\theta)\cos\phi\right].
\end{equation}
Here, $\sigma_0$ represents the unpolarized cross section, $\theta$
and $\phi$ denote the polar and azimuthal angles of the scattered
proton as shown in Fig.~\ref{fig:illustration_coordinate};
$\mathcal{P}_y$ corresponds to the normal polarization component of
the incident proton to be determined; $A_N(\theta)$ is the analyzing
power, which depends on the scattering angle, momentum of the incident
proton, and the target material. It has been measured extensively for
both the $pp$ and $pC$ scattering
processes~\cite{PhysRev.148.1289,PhysRev.163.1470, PhysRev.95.1348,
ALBROW1970445, PhysRevC.24.1778, vonPrzewoski:1998ye,
PhysRevLett.41.384, PhysRevD.21.580, PhysRevD.40.35, PhysRev.105.288,
GREENIAUS1979308}.

Proton polarimeters utilizing $pp$ or $p\textrm{C}$ scattering have
been widely employed, primarily as dedicated detectors for measuring
initial-state proton beam polarization or final-state protons within
limited acceptance. For a large acceptance general purpose
spectrometer, a novel technique has been developed to measure proton
polarization based on the BESIII detector
configuration~\cite{c642-1lzb}. Using polarized protons from hyperon
decays, this technique has been validated and proves that the BESIII
detector can work as a proton polarimeter. In this Letter, we present
the first determination of the relative phase between the proton
psionic FFs via the measurement of final-state proton polarization in
the $e^+e^- \rightarrow J/\psi \rightarrow p\bar{p}$ process using this
novel technique~\cite{c642-1lzb}.

The BESIII detector records symmetric $e^+e^-$ collisions provided by
the BEPCII storage ring operating in the c.m.\ energy range
$\sqrt{s}=1.84$ to $4.95$~GeV. BESIII has collected $10\times10^{9}$
$J/\psi$ events~\cite{BESIII:2021cxx} including processes such as
$J/\psi\to p\bar p$ with a branching fraction
$2.12(3)\times10^{-3}$~\cite{PhysRevD.110.030001}.  The BESIII
detector consists of a helium-based multilayer drift chamber~(MDC), a
plastic scintillator time-of-flight system~(TOF), and a CsI(Tl)
electromagnetic calorimeter~(EMC), which are all enclosed in a
superconducting solenoidal magnet providing a 1.0~T magnetic
field. The momentum resolution of charged particle at $1~{\rm GeV}/c$
is $0.5\%$. A detailed description of the detector and its performance
can be found in Refs.~\cite{ABLIKIM2010345,CAO2020163053}.

At BESIII, a layer of cooling fluid of high purity mineral oil with
a thickness of 0.8~mm circulates between the inner and outer
beryllium cylinders of the beam pipe. The inner radius of the oil layer
is 32.3~mm~\cite{10.1088/1674-1137/ad3dde,
ABLIKIM2010345,PhysRevLett.127.012003}. Mineral oil, composed
primarily of carbon and hydrogen, is an ideal target material for
nucleon polarization measurements. The tracking detector has an inner
wall made of carbon fiber with a thickness of 1.2 mm. It contains
mainly carbon, hydrogen, and a small fraction of oxygen, which is
also a good target candidate. These supporting components of the
detector were optimized to be as thin as
possible to improve detector performance. BESIII has demonstrated
excellent reconstruction capability for scattering events at these
materials, as shown by previous hyperon--nucleon scattering studies
that successfully employed secondary interactions in the beam
pipe~\cite{BESIII:2023clq,BESIII:2024geh,PhysRevC.109.L052201}. While
carbon is an effective material for analyzing proton polarization, its
use in $p\textrm{C}$ scattering at BESIII presents technical
challenges for background suppression, as the low-momentum recoil
carbon nuclei cannot be detected. Therefore, this study uses only
hydrogen as an analyzer, where both the target proton
and the scattered proton are detected, overcoming this limitation.

In this analysis, the candidate events are
$e^{+}e^{-}p_{\rm{pipe}}\rightarrow J/\psi p_{\rm{pipe}} \rightarrow
p\bar{p}p_{\rm{pipe}}$ and contain three charged particles ($p,
\bar{p}, p_{\rm{pipe}}$), where $p_{\rm{pipe}}$ is a proton knocked
out from the target layer, as illustrated in
Fig.~\ref{fig:illustration_coordinate}. Charged tracks are
reconstructed in the MDC within the angular region
$\vert\!\cos\theta\vert< 0.93$, where $\theta$ is the polar angle with
respect to the $z$ axis in the laboratory frame. Particle
identification~(PID) for charged tracks combines measurements of the
energy deposited in the MDC~(d$E$/d$x$) and the flight time in the TOF
to form likelihoods $\mathcal{L}(h)~(h=p,K,\pi)$ for each hadron $h$
hypothesis. Tracks are identified as protons when the proton
hypothesis has the greatest likelihood
($\mathcal{L}(p)>\mathcal{L}(K)$ and
$\mathcal{L}(p)>\mathcal{L}(\pi)$).

\begin{figure}
\includegraphics[width=0.45\textwidth]{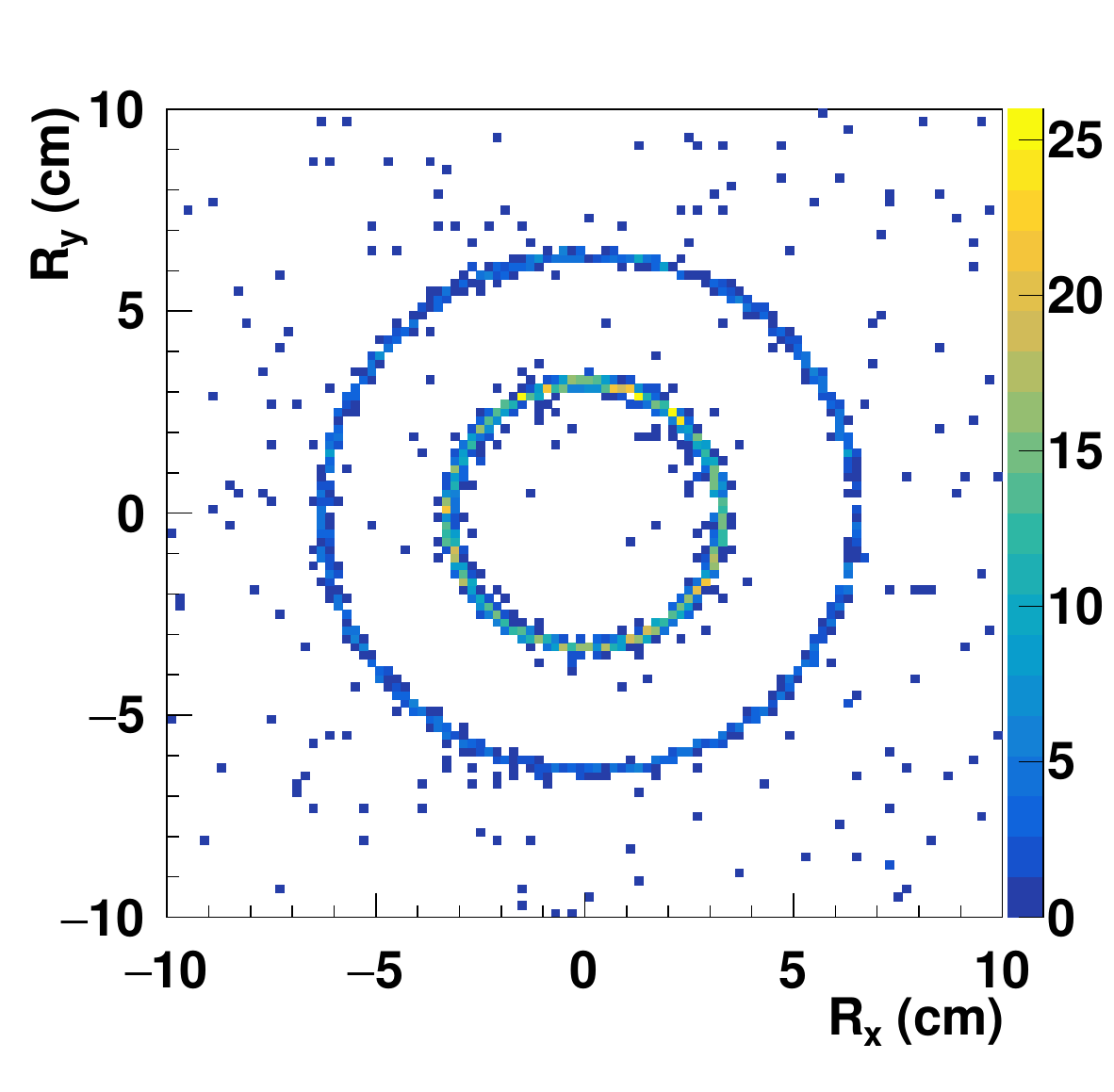}
\caption{\label{fig:Rxy_pp} {Distribution of $pp$ scattering vertices at the beam pipe and the inner wall of the MDC.}
}
\end{figure}

A vertex fit is performed to the two final-state protons, and the
vertex distribution of $pp$ candidate events is displayed in
Fig.~\ref{fig:Rxy_pp}, where the beam pipe structure and the MDC inner
wall are clearly resolved. The vertex is required to be at a distance $R_{xy}$ from the axis of the beam pipe of  $3.0 < R_{xy}
< 3.5$~cm, or $6.0 < R_{xy} < 6.8$~cm, corresponding to a $\pm
4\sigma$ window. Since scattering occurs
upstream of the tracking detectors, the incident proton momentum is
determined through recoil kinematics: $p_p = p_{e^+} + p_{e^-} -
p_{\bar{p}}$. Here, $p_{e^{\pm}}$ represents the beam four-momenta. We
calculate the target proton momentum ($p_{\rm{target}}$) using
four-momentum conservation between the two final-state protons and the
incident proton, and require the momentum of the target proton
$|p_{\rm{target}}| <  50$~MeV/$c$. This requirement suppresses a
type of background collision where the incident proton hits a proton
in a nucleus like carbon, beryllium, etc. In this background
collision, the target proton is not at rest as for a free hydrogen target,
but has a sizable Fermi momentum. This collision is referred to as a
``quasi-free nuclear collision" in this context.

\begin{figure}[t!]
\centering
\includegraphics[width=0.45\textwidth]{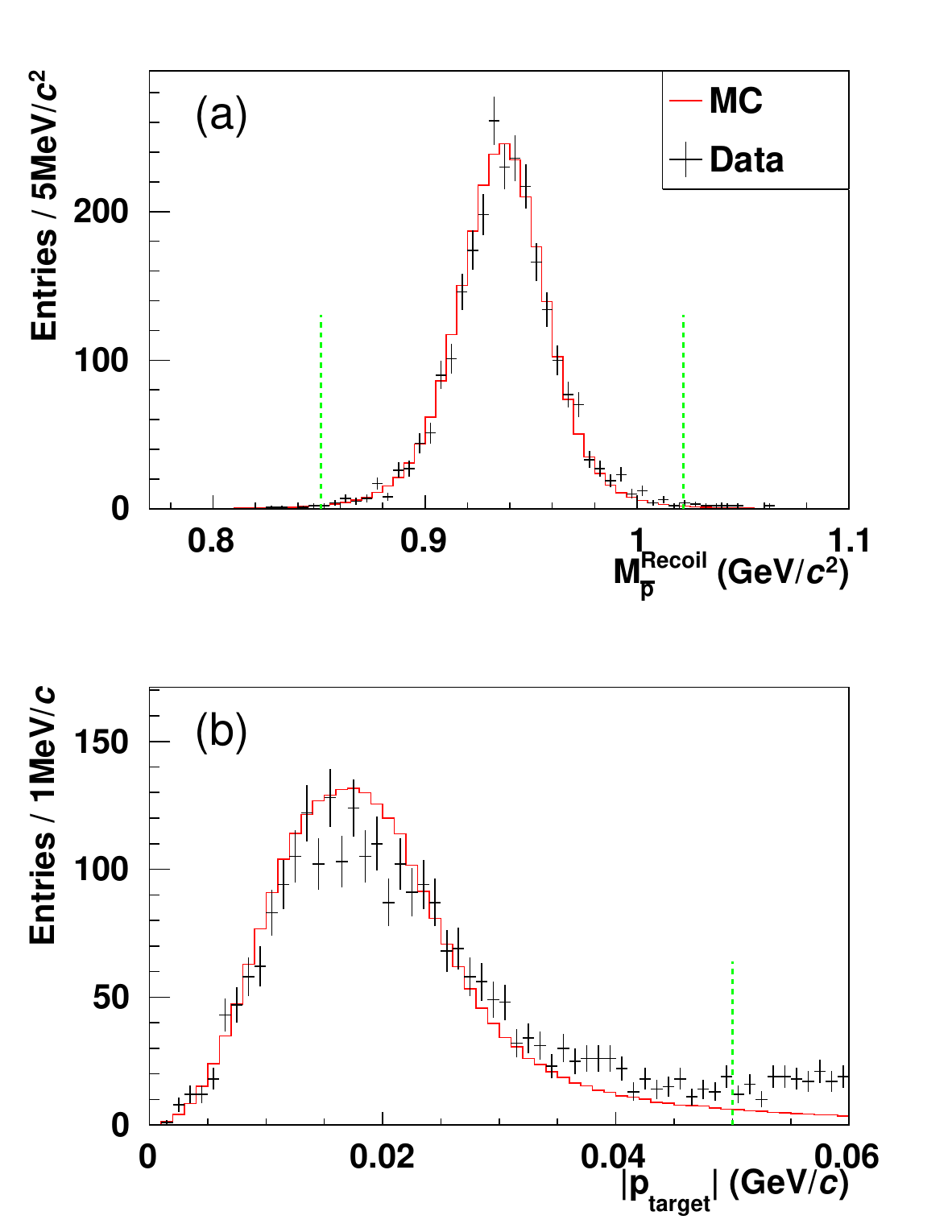}
\caption{Distributions of $pp$ in elastic scattering signal events:  mass $M_{\bar p}^\text{Recoil}$ recoiling against the antiproton (top), and inferred momentum $|p_\text{target}|$  of the target proton. The
red histograms are signal MC, the black points with error bars are the
data, and the green vertical lines indicate the requirements applied.}
\label{fig:Sig_pp}
\end{figure}

Figure~\ref{fig:Sig_pp} shows the surviving elastic $pp$ scattering
candidates. The experimental data are compared with Monte Carlo (MC)
simulations, with the selection requirements indicated by dashed green
lines. The signal MC event generation is carried out in two
stages. First, the decay process $J/\psi\rightarrow p\bar{p}$ is
generated using the standard {\sc EvtGen}
generator~\cite{LANGE2001152, PingRong-Gang_2008}. Subsequently, the
produced proton is transported to the beam pipe, where it interacts
with a stationary target proton. Finally, the three resulting
final-state particles (the initial $\bar{p}$, and two protons
originating from the beam pipe) are simulated by the {\sc Geant4}
BESIII software. Figure~\ref{fig:Sig_pp} (a) shows the recoil mass
distribution of $\bar{p}$, where candidates must satisfy 0.85 $<
M^{\text{Recoil}}_{\bar{p}} <$ 1.02
GeV/$c^2$. Figure~\ref{fig:Sig_pp}(b) shows the momentum distribution
of the target proton. The number of surviving $pp$ scattering signal
events is 2556, with a detector efficiency of 15.2\%.

\begin{figure}[htbp]
\centering
\includegraphics[width=0.45\textwidth]{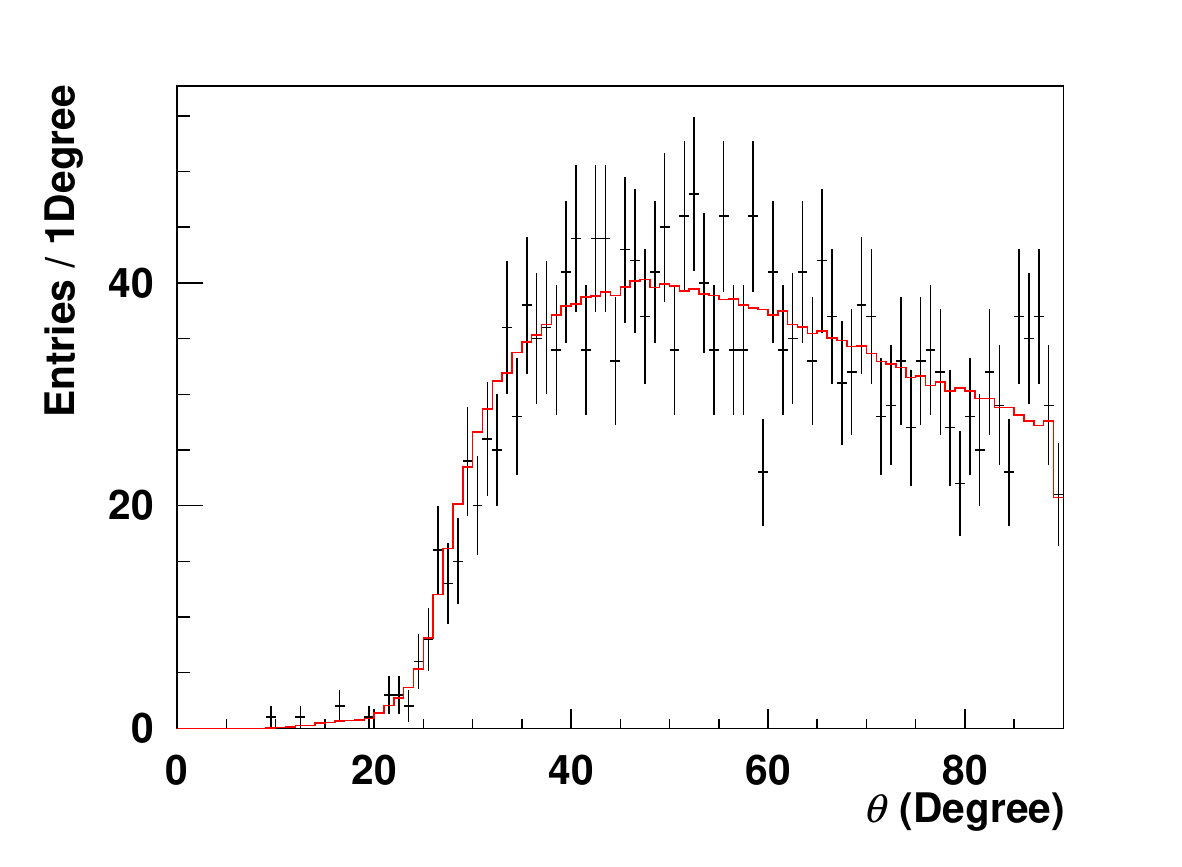}
\caption{Polar angle distribution of the scattered proton in the $pp$ c.m. frame. The red histogram is signal MC, the black points with error bars are the data.}
\label{fig:An_pp_theta_bin}
\end{figure}

Figure~\ref{fig:An_pp_theta_bin} shows the polar angle ($\theta$)
distribution of the scattered protons. For each event, the expected
analyzing power $A_N(\theta)$ can be obtained from the SAID
database~\cite{PhysRevC.62.034005, *PhysRevC.76.025209, *GWDAC}. The
normal polarization of protons, $\mathcal{P}_y$, can be obtained from
the azimuthal angle distribution of $pp$ scattering. Based on
Eq.~\ref{eq:py_angular_dependent}, $\mathcal{P}_y$ depends on the polar angle
$\theta_p$ and the relative phase $\Delta\Phi$. In this
analysis we use an unbinned maximum likelihood fit based on the
angular distribution \begin{equation} \begin{aligned}
&\mathcal{W}(\theta_{p}, \phi; \sin\Delta\Phi) \\ & \propto
(1+\alpha_{J/\psi}\cos^{2}\theta_{p})(1+\mathcal{P}_yA_N(\theta)\cos\phi).
\label{eq:maximum_fit_function_1} \end{aligned} \end{equation}
The likelihood function is defined as
\begin{equation}
\begin{aligned} 
\mathcal{L}  = \prod_{i=1}^{N}\mathcal{CW}(\theta_{p}^{i}, \phi^{i}; \sin\Delta\Phi)\epsilon(\theta_{p}^{i}, \phi^{i}).
\label{eq:maximum_fit_function_3}
\end{aligned}
\end{equation}
Here, $N$ is the number of events. The detection efficiency is denoted
by $\epsilon(\theta_{p}^{i}, \phi^{i})$. The normalization factor
$\mathcal{C}^{-1} = \frac{1}{N_{\rm MC}}\prod_{j=1}^{N_{\rm
MC}}\mathcal{W}(\theta_{p}^{j},
\phi^{j};\sin\Delta\Phi)\epsilon(\theta_{p}^{j}, \phi^{j})$ is
estimated with the $N_{\rm MC}$ events generated according to the
phase space model of $J/\psi\rightarrow p\bar{p}$ with unpolarized
scattering ($\sin\Delta\Phi=0$). These MC events are subject to the
same selection criteria as the experimental data. To ensure sufficient
statistical precision, we generate approximately 500 times more MC
events than the accepted data sample. The background level is about
0.5\% and is considered in the systematic uncertainty evaluation. For
the minimization procedure, we employ {\sc
MINUIT}~\cite{James:1975dr} from the CERN library to minimize
the objective function $-\ln\mathcal{L}$.

\begin{figure}[htbp]
\centering
\includegraphics[width=0.51\textwidth]{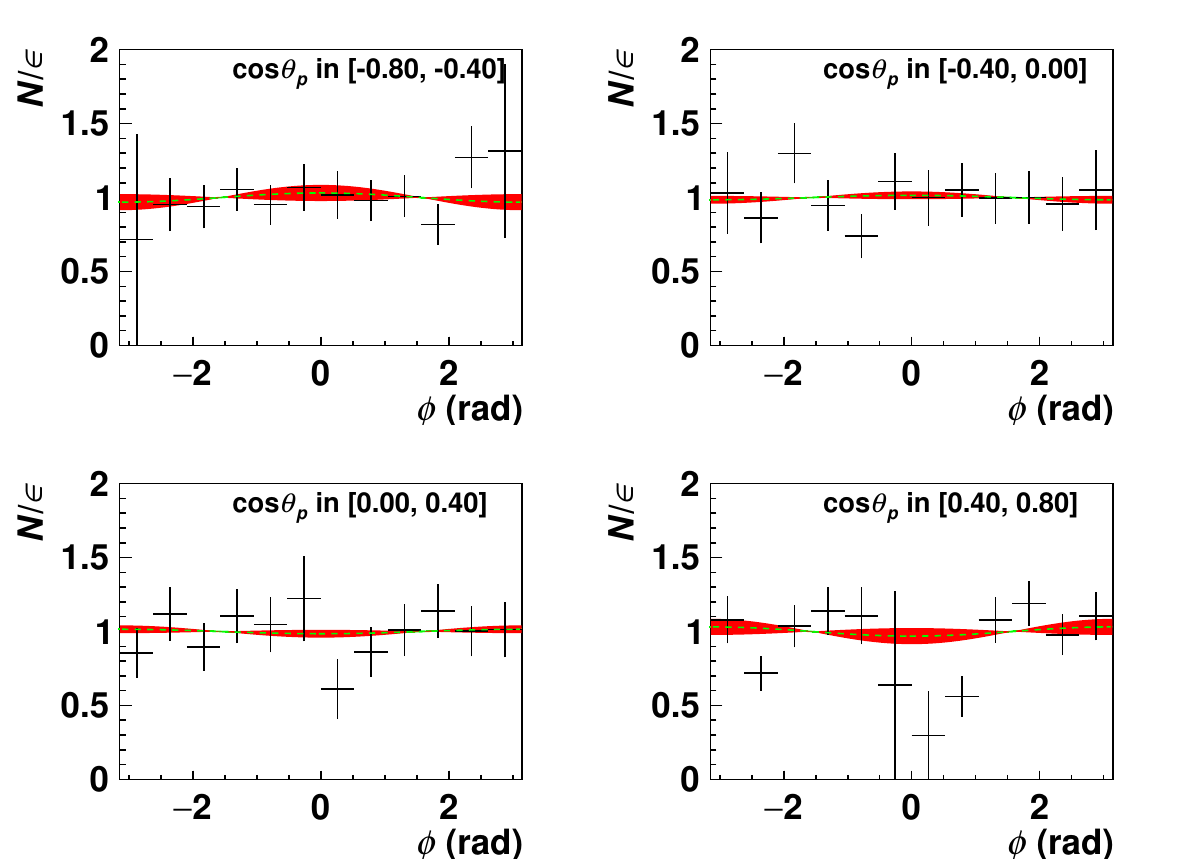}
\caption{Distributions of the azimuthal angle $\phi$ of the scattered proton
in each $\cos(\theta_p)$ bin. The black points with error bars are the
experimental data. The red bands with green dashed lines in the middle
are the unbinned fit. $N/\epsilon$ denotes the normalized efficiency
corrected yield.}
\label{fig:SSA_Fit_data}
\end{figure}

In the maximum likelihood fit, the parameter $\alpha_{J/\psi}$ is
fixed at 0.595 based on the previous BESIII
measurement~\cite{BESIII:2012ion}. The fit yields a value of
$\sin\Delta\Phi=-0.20\pm0.34$. To illustrate the fit quality, we
present the results binned in four $\cos\theta_p$ intervals in
Fig.~\ref{fig:SSA_Fit_data}, where the black data points with error
bars are the experimental measurements and the red bands show the
unbinned fit projections. A comprehensive validation is performed
using pseudo experiments generated with an input
$\sin\Delta\Phi=-0.84$. Applying the full {\sc
Geant4}~\cite{AGOSTINELLI2003250} simulation, and the same unbinned
maximum-likelihood fitting procedure, we obtain $\sin\Delta\Phi=-0.82
\pm 0.02$, where the excellent agreement with the input parameter
confirms the reliability of our analysis method.

The proton spin undergoes precession in the BESIII solenoid magnetic
field, resulting in a measurable difference between the polarization
states at the production point and at the scattering target
(positioned at radii of either 3.2 cm or 6.3 cm). We simulate this
spin precession using {\sc Geant4}~\cite{AGOSTINELLI2003250} and
account for its effects through a rotation of the coordinate system rotation.

The systematic uncertainties in this analysis can be categorized into
three components: 1) uncertainties from event selection, including
background estimation, tracking and PID; 2) fit procedure
uncertainties; 3) uncertainties from the analyzing power. In the event
selection, the $\bar{p}$ is detected, and its recoil side is required
to have a proton mass with a mass window of 4$\sigma$ determined from
the MC simulation. The requirement is changed between 3$\sigma$ and
5$\sigma$. With a Barlow test~\cite{Barlow:2002yb}, the most
significant difference on the fit results is taken as the systematic
uncertainty. The requirement of $p < 50$ MeV/$c$ for target protons is
varied between 40 MeV/$c$ and 70 MeV/$c$, with the most significant
difference in fit results taken as the systematic uncertainty. Using a
$J/\psi\rightarrow p\bar{p}\pi^+\pi^-$ control sample, the $p$ and
$\bar{p}$ tracking and PID efficiencies for both data and MC
simulations are determined.  The data/MC efficiency ratios for $p$ and
$\bar{p}$ are applied as correction factors to the MC simulation.  To
reduce the impact of statistical fluctuations, the fit with the
corrected MC sample is performed 100 times with correction factors
varied within their $\pm1\sigma$ uncertainties. The deviation between
the mean corrected and nominal fit results is taken as the systematic
uncertainty. The primary vertex in the MC simulation is shifted by
$\pm1\sigma$, with the largest fit variation treated as the
systematic uncertainty. The polar angle distribution of $pp$
scattering is modeled using a data-driven approach. We generate
alternate MC samples with different polynomial orders for the angular
distribution and use the largest difference of the result as the systematic
uncertainty.

The fit procedure uncertainty is evaluated using MC simulation. It is
taken as the difference between the input and output values of the
fit. The analyzing power for $pp$ scattering is well measured in the
relevant energy regime~\cite{GWDAC}. We assign a 2\% uncertainty to
account for potential variations in the reported values. Quasi-free
collisions in the target material may further affect the analyzing
power. MC simulations indicate that these collisions could constitute
up to 7.7\% of events. Conservatively, we assume these quasi-free
collisions having zero analyzing power, and the difference in the fit
results is taken as the systematic uncertainty. The potential impact
from BEPCII beam polarization is also evaluated. Due to synchrotron
radiation, electron and positron beams may acquire sizable transverse
polarization. In Ref.~\cite{PhysRevD.110.014035}, it is deduced that
the transverse polarization of $\Lambda$ (and other baryons
with spin 1/2) will have an extra dependence on $\cos 2\phi$ and $\sin
2\phi$, with $\phi$ as the azimuthal angle of proton momentum in the
laboratory frame. In this work, the normal polarization of the proton
is measured over the full range of $\phi$, in which the integrals of
extra terms with $\cos2\phi$ and $\sin2\phi$ are zero. The impact from
beam transverse polarization is naturally eliminated.  The combined
systematic uncertainty on the $\sin\Delta\Phi$ value is determined to
be $\pm0.113$. The details are listed in
Table~\ref{table:sys_err_validation_helicitiy_angle}.

\begin{table}[h!]
\centering
\begin{tabular}{ l | c  }
\hline
 Source & $\sigma_{\sin(\Delta\Phi)}^p$    \\\hline
 Mass window & 0.040      \\
 Momentum requirement & 0.049     \\
 Efficiency correction & 0.055   \\
 Primary vertex  &  0.013   \\
 Polar angle distribution & 0.021   \\
 Analyzing power & 0.009     \\
 Fit method  &  0.010   \\ 
 Quasi-free collision &0.070\\  
 \hline
 Sum          &0.113     \\\hline  
 
\end{tabular}
\caption{Systematic uncertainties in the proton polarization measurement.}
\label{table:sys_err_validation_helicitiy_angle}
\end{table}

In summary, a novel technique enabling the BESIII detector to function
as a proton polarimeter is used to measure the proton polarization in
$e^+e^-\rightarrow J/\psi \rightarrow p\bar{p}$, allowing the sine of
the relative phase between the proton psionic FFs to be determined for
the first time,
$\sin\Delta\Phi=-0.20\pm0.34_{\textrm{stat}}\pm0.11_{\textrm{syst}}$. The
result yields two solutions for $\Delta\Phi$, which can only be
decided by measuring an additional polarization observable, such as
the sideways polarization $\mathcal{P}_x$ (with a polarized beam) or
measuring the antiproton polarization.  Future experiments in
high-luminosity electron-positron colliders, such as the Super
Tau-Charm Facility (STCF)~\cite{Achasov:2023gey}, will resolve this
ambiguity and allow a clear determination of $\Delta\Phi$.

\begin{figure}[htbp]
\centering
\includegraphics[width=0.48\textwidth]{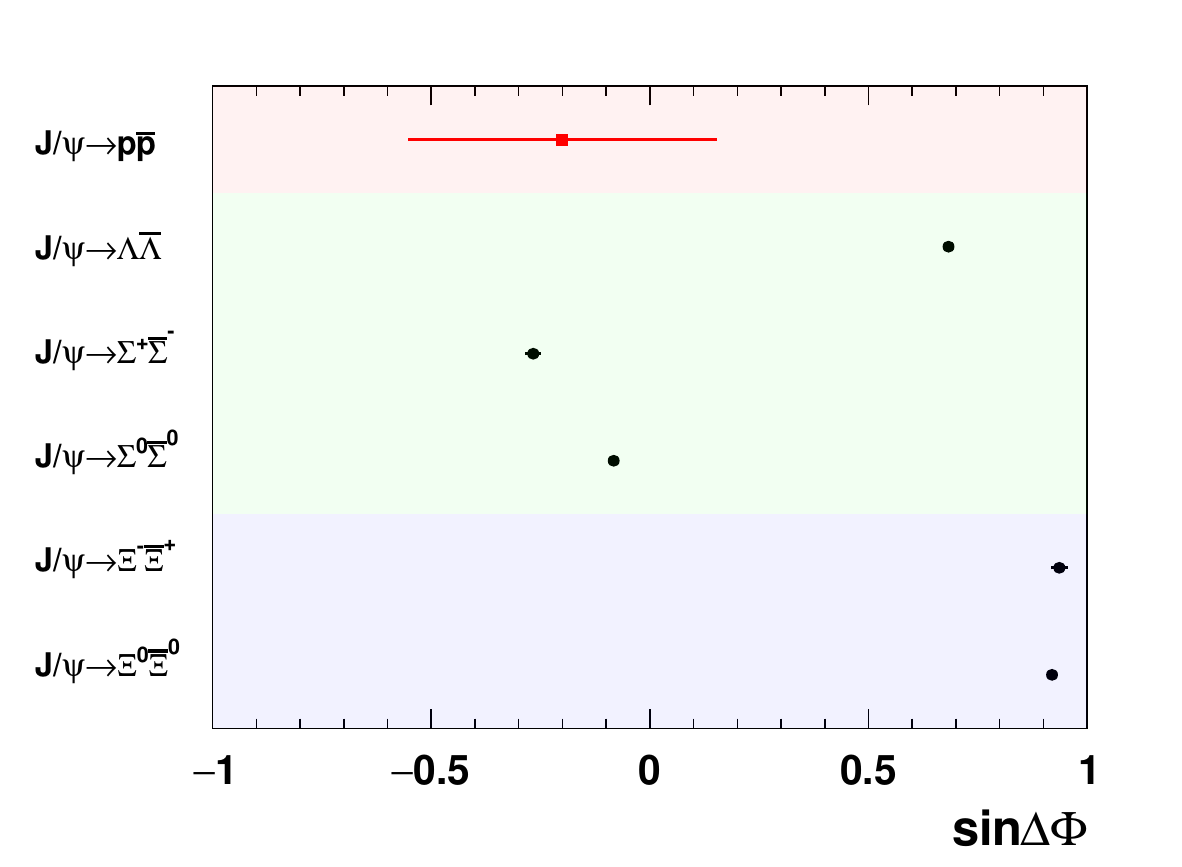}
\caption{Relative phase of form factors for baryon octet states. The
rectangle point in red for the proton is from this work. The other five
round points are from Refs.~\cite{PhysRevLett.129.131801,
PhysRevLett.125.052004, PhysRevLett.133.101902, PhysRevD.108.L031106,
nature_BESIII}. The six baryons are categorized into three color bands
according to the strangeness number for visualization purpose.}
\label{fig:dphi_octet}
\end{figure}

As a first-ever measurement for the relative phase of proton psionic
form factors, a comparison with the measurements of other baryon
octets can provide a meaningful test on SU(3)
symmetry~\cite{PhysRevD.77.056001}. Fig.~\ref{fig:dphi_octet} shows
the $\sin\Delta\Phi$ of six baryon octet states from $J/\psi$ decay,
including single-strange baryons
$\Lambda$~\cite{PhysRevLett.129.131801},
$\Sigma^+$~\cite{PhysRevLett.125.052004},
$\Sigma^0$~\cite{PhysRevLett.133.101902}, and double-strange baryons
$\Xi^-$~\cite{nature_BESIII}, $\Xi^0$~\cite{PhysRevD.108.L031106}. The
relative phase of the proton is close to that of $\Sigma^{+/0}$ and is
smaller than that of $\Lambda$ and $\Xi^{-/0}$. The two unmeasured states
in the baryon octet sector are $\Sigma^-$ and the neutron. The relative
phase for $\Sigma^-$ can be measured in the current BESIII
experiment. Future experiments with better neutron detection
performance may have the opportunity to measure the relative phase of
neutron psionic form factors utilizing the same method as this work by
neutron scattering off hydrogen or the carbon nucleus.

\textbf{Acknowledgement}

The BESIII Collaboration thanks the staff of BEPCII (https://cstr.cn/31109.02.BEPC) and the IHEP computing center for their strong support. This work is supported in part by National Key R\&D Program of China under Contracts Nos. 2023YFA1606800, 2023YFA1606704, 2025YFA1613900, 2023YFA1606000; National Natural Science Foundation of China (NSFC) under Contracts Nos. 12275320, 11975278, 11635010, 11935015, 11935016, 11935018, 12025502, 12035009, 12035013, 12061131003, 12192260, 12192261, 12192262, 12192263, 12192264, 12192265, 12221005, 12225509, 12235017, 12342502, 12361141819; the Chinese Academy of Sciences (CAS) Large-Scale Scientific Facility Program; the Strategic Priority Research Program of Chinese Academy of Sciences under Contract No. XDA0480600; CAS under Contract No. YSBR-101; 100 Talents Program of CAS; The Institute of Nuclear and Particle Physics (INPAC) and Shanghai Key Laboratory for Particle Physics and Cosmology; ERC under Contract No. 758462; Beijing Municipal Natural
Science Foundation under Contracts No. JQ22002; German Research Foundation DFG under Contract No. FOR5327; Istituto Nazionale di Fisica Nucleare, Italy; Knut and Alice Wallenberg Foundation under Contracts Nos. 2021.0174, 2021.0299, 2023.0315; Ministry of Development of Turkey under Contract No. DPT2006K-120470; National Research Foundation of Korea under Contract No. NRF-2022R1A2C1092335; National Science and Technology fund of Mongolia; Polish National Science Centre under Contract No. 2024/53/B/ST2/00975; STFC (United Kingdom); Swedish Research Council under Contract No. 2019.04595; U. S. Department of Energy under Contract No. DE-FG02-05ER41374.



\bibliography{polarimeter_update}

\end{document}